# Dynamic interactions between deformable drops in the Hele-Shaw geometry

Derek Y. C. Chan[1,2*], Evert Klaseboer[3] and Rogerio Manica[3]

[1]Particulate Fluids Processing Centre, Department of Mathematics and Statistics,
The University of Melbourne, Australia 3010
[2]Department of Mathematics, National University of Singapore, Singapore 117543
[3]Institute of High Performance Computing, 1 Fusionopolis Way, #16-16 Connexis,
Singapore 138632

[*]Email: D.Chan@unimelb.edu.au

A model has been developed to describe the collision and possible coalescence of two driven deformable drops in the Hele-Shaw cell. The interdependence between hydrodynamic effects and interfacial deformations is characterised by a film capillary number: $Ca_f = (\mu v/\sigma)(R_o/h_o)^{3/2}$ as revealed by an analytic perturbation solution of the governing equations for a system with continuous phase viscosity $\mu$, interfacial tension $\sigma$, drop radius $R_o$, characteristic relative velocity $v$ and separation $h_o$ between the drops. Numerical solutions of the model demonstrate the importance of the full dynamic history of the interacting drops in determining stability or coalescence. The geometry of the Hele-Shaw cell allows for the possibility of using the model to infer the time dependent force between colliding drops by measuring their separation.

## 1. Introduction

There are a number of complementary experimental and theoretical approaches to quantify the interaction dynamics between deformable drops and bubbles driven together or separated by applied external fields. For example, the four-roll mill has been used extensively as a well-defined experimental and theoretical system to study interaction between polymeric drops under external flow fields[1,2]. The time-dependent deformations and drainage of the intervening non-polar liquid film between drops of polar fluids attached to the ends of thin approaching capillaries have been measured[3] and modelled[4] with good quantitative agreement. Interferometric techniques have been used to measure, with sub-nanometre precision, complex dynamic deformations of a mercury drop moving in water towards and away from a mica surface[5,6] and the results can be explained in terms of the combined effects of hydrodynamics, electrical double layer forces and interfacial deformations[7,8]. The atomic force microscope has been used to make direct measurements of dynamic forces between approaching and receding emulsion drops[9] or micro-bubbles[10] in water to elucidate the effects of hydrodynamic boundary conditions, drainage of nanometer thick films and interfacial deformations.

Recently, the coalescence of two water drops in hexadecane has been studied in a microfluidic cell in which the drops can be driven together or separated by designing the cell geometry to control flow rates[11] and a Hele-Shaw type model[12] has been proposed to explain the observed coalescence between separating drops. However, the treatment of this model is flawed because of an incorrect sign in the governing equation for the film thickness and an incorrect boundary condition has been imposed when considering flow and deformation in the interaction zone which resulted in an unphysical cusp in the film thickness at the point of closest approach between the interacting drops.

In this paper, we consider the dynamic interaction between two drops in a Hele-Shaw cell geometry in the limit where viscous stresses on the drops are small compared to their interfacial tension so that deformations are small compared to the drop dimensions. A complete analysis requires consideration of flow and deformation in the small inner interaction zone between the two drops coupled with a consistent account of global drop deformation outside the interaction zone that provides the outer correct boundary condition for the inner problem. An analytic perturbation solution is also obtained and its range of validity is quantified by comparison with the full numerical solution of the governing equations. This paper complements a similar study of the interaction between axi-symmetric drops[13].

The structure of this paper is as follows. In section 2, we derive the connection between the local ('inner') and global ('outer') deformations of a drop in a Hele-Shaw cell that is subjected to a localized weak force density around the osculation region or interaction zone between two interacting drops. Hydrodynamic effects and surface forces that are only important in the small inner interaction zone are treated by a lubrication approximation in section 3. The outer solution which accounts for volume conservation of the drop provides the required boundary condition to completely determine the inner solution. Such an approach has been used before in analyzing buoyancy driven interaction of deformable drops[14] and in analyzing equilibrium surface force measurements involving deformable drops[15,16]. In section 4 we present an analytical perturbation solution of the governing equations and compare the results with that of Lai *et al*. Numerical solutions of the governing equations for approaching and separating drops are compared with predictions of the perturbation theory in section 5. The paper closes with a discussion in which we propose exploiting the analytic form of the perturbation solution to devise a novel method to measure the magnitude of dynamic forces between interacting drops from simply observing the drop separation.

**2. Deformations of a Hele-Shaw drop**

Consider two proximal drops in a Hele-Shaw cell in the *xy*-plane (see Fig. 1) with thickness or depth (2*b*) in the *z* direction. The interaction between the drops is described by a pressure profile $p(x,t)$ localised in the osculation region between the drops which deform from the unperturbed circular shape of radius $R_o$, with $R_o \gg b$. This pressure profile can be due to hydrodynamic interaction as well as surface forces. The position dependent separation between the drops is $2h(x,t)$. The deformed boundary of the drops $y(x,t)$ with interfacial tension $\sigma$ is given by the augmented Young-Laplace equation that relates the pressure jump across the interface of the drop to the mean curvature

$$\sigma\left(\frac{1}{R_1} + \frac{1}{R_2}\right) = \Delta P - p \qquad (2.1)$$

where $\Delta P$ is the Laplace pressure of the drop. In the Hele-Shaw cell geometry, the Young-Laplace equation is approximated by

$$\sigma\left(\frac{1}{b} - \frac{\partial}{\partial x}\left[\frac{\partial y/\partial x}{\left[1 + (\partial y/\partial x)^2\right]^{1/2}}\right]\right) = \frac{\sigma}{b} + \frac{\sigma}{R} - p. \qquad (2.2)$$

The status of this depth-averaged equation in the context of describing interfacial deformations in the Hele-Shaw cell has been discussed in detail by Homsy[17,18]. Here $R$ can be regarded as the Lagrange multiplier that ensures the drops deform in the *xy*-plane under a constant volume constraint.

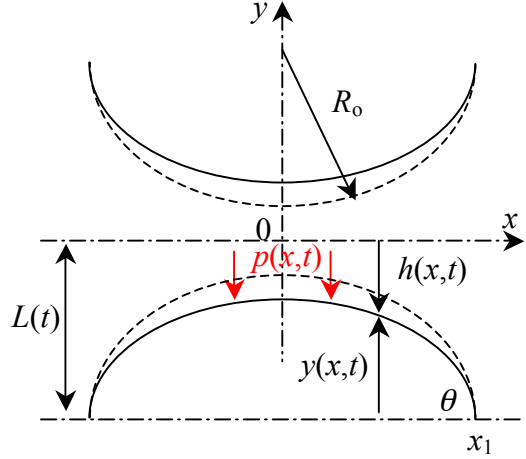

**Fig. 1** Schematic representation of the interface (—) of two identical interacting drops in a Hele-Shaw cell with film thickness ($2h$). The drops are deformed under the action of a localized normal pressure distribution $p$ relative to the undeformed drops of radius $R_o$ (- -).

The interacting drops are taken to be symmetric about $x = 0$ so we only need to consider $x > 0$. If the pressure $p(x,t)$ arising from the drop-drop interaction is localised around $x = 0$ over a small range $x_p \ll R_o$, eqn (2.2) has two distinct regions of behaviour if the deformation is small on the scale of the drop. We call the region $0 < x < x_p$ the *inner* region where $|\partial y/\partial x| \ll 1$, with an *inner* solution of eqn (2.2) that obeys

$$\frac{\partial^2 y_{in}}{\partial x^2} = -\frac{1}{R} + \frac{p}{\sigma} \ . \tag{2.3}$$

This can be integrated with the symmetry condition: $\partial y_{in}/\partial x = 0$, at $x = 0$ to give the outer asymptotic behaviour valid for $x_p \ll x \ll R$

$$y_{in}(x,t) \rightarrow \{y(0) - k(t)\} - \frac{x^2}{2R} + f(t)\, x, \qquad x \gg x_p \tag{2.4}$$

where

$$f(t) \equiv \frac{1}{\sigma} \int_0^\infty p(x,t)\, dx \tag{2.5}$$

$$k(t) \equiv \frac{1}{\sigma} \int_0^\infty x\, p(x,t)\, dx .$$

where the force acting on the drop is $4b\sigma f(t)$. Small deformations correspond to the dimensionless force $f \ll 1$. The linear and quadratic behaviour in $x$ will be matched to the outer solution which will also determine the value of $y(0)$.

In the *outer* region, $x > fR > x_p$, the *outer* solution of eqn (2.2) obeys

$$\frac{\partial y_{out}/\partial x}{\left[1 + (\partial y_{out}/\partial x)^2\right]^{1/2}} = -\frac{x}{R} + f \tag{2.6}$$

where we assume the drop subtends an acute contact angle $\theta \leq \pi/2$ at $x_1$ (see Fig. 1). Eqn (2.6) can be integrated to give the outer shape of the drop

$$y_{out}(x,t) = R\sqrt{1 - (f - \frac{x}{R})^2} - R\sqrt{1 - (f - \frac{x_1}{R})^2} \quad . \tag{2.7}$$

For $f \ll 1$, the inner asymptotic behaviour, $x \ll R$, of $y_{out}(x)$ is, to linear order in $f$,

$$y_{out}(x,t) \rightarrow \left\{ R\left[1 - \sqrt{1 - (x_1/R)^2}\right] - \frac{f\,x_1}{\sqrt{1 - (x_1/R)^2}} \right\} - \frac{x^2}{2R} + f\,x \tag{2.8}$$

As expected, the inner asymptotic behaviour of $y_{out}(x,t)$ in eqn (2.8) and the outer asymptotic behaviour of $y_{in}(x,t)$ in eqn (2.4) have matching functional forms. In particular, the constant terms in braces in eqns (2.4) and (2.8) are the same and this determines $y(0)$ in terms of $R$ and $f$. The remaining task is to use the constant volume constraint to determine the relationship between the radius of curvature $R$ of the deformed drop and the dimensionless force $f$.

The volume of the undeformed drop with radius of curvature $R_o$ and contact angle $\theta_o$ is (see Fig. 1):

$$V_o = 2b\,R_o^2 \{\theta_o - \sin\theta_o \cos\theta_o\} \tag{2.9}$$

To linear order in $f$, the volume of deformed drop $V_{def}$ with radius of curvature $R$ can be found by integrating only the outer solution $y_{out}(x)$ given by eqn (2.7) since the volume from the inner solution $y_{in}(x)$ does not contribute to linear order in $f$. Thus

$$V_{def} = 4b \int_0^{x_1} y_{out}(x,t)\,dx$$

$$= 2bR \left\{ Rf\sqrt{1-f^2} + R\left[\arcsin(f) - \arcsin(f - \frac{x_1}{R})\right] \right.$$

$$\left. - (Rf + x_1)\sqrt{1 - (f - \frac{x_1}{R})^2} \right\} \tag{2.10}$$

By putting $R \equiv R_o(1+\rho)$, and imposing the constant volume constraint $V_{def} = V_o$ we find, after some algebra, that to linear order in $f$ and $\rho$

$$\rho = -\left\{\frac{1-\cos\theta_o}{\sin\theta_o - \theta_o\cos\theta_o}\right\} f. \tag{2.11}$$

With this result, the outer asymptotic form of the inner solution $y_{in}(x)$ in eqn (2.4), can be cast into a physically perspicuous form:

$$y_{in}(x,t) \approx R_o(1-\cos\theta_o) - f(t)R_o\left\{\frac{2-2\cos\theta_o - \theta_o\sin\theta_o}{\sin\theta_o - \theta_o\cos\theta_o}\right\} - \frac{x^2}{2R} + f(t)x \tag{2.12}$$

where the first term on the rhs is the height of the undeformed drop and the second represents a correction to the drop height due to the applied force $f$. As the term in braces is positive, the effect of approaching drops will give rise to a compressive force $f > 0$, so the drop height will be reduced as expected as a result of the interaction. One the other hand, for separating drops for which $f < 0$, this correction term that is proportional to $f$ will cause the drop interfaces to deform towards each other to give a thinner film than expected. Note that because of volume conservation, the global deformation of the drop has an influence on the form of drop deformation in the inner region.

### 3. Film drainage between drops

In view of the geometric relation: $L(t) = h(x,t) + y(x,t)$ (see Fig. 1) and eqn (2.3), the half-thickness, $h(x,t)$, of the film between the drops obeys the equation

$$\sigma\frac{\partial^2 h}{\partial x^2} = \frac{\sigma}{R} - p \tag{3.1}$$

in the inner interaction zone.

The time evolution of the film can be treated by the usual lubrication theory for the dominant *x*-component of the fluid velocity $u(y)$ which in Stokes flow obeys: $\partial p/\partial x = \mu\, d^2u/dy^2$, where $\mu$ is the viscosity of the continuous phase. Applying the no-slip boundary condition $u(\pm h) = 0$ at the drop surfaces gives the solution: $u(y) = (1/2\mu)(\partial p/\partial x)(y^2 - h^2)$. Integrating the continuity equation with the kinematic condition at the drop surfaces give the Stokes-Reynolds film drainage equation

$$\frac{\partial h}{\partial t} = \frac{1}{3\mu}\frac{\partial}{\partial x}\left(h^3 \frac{\partial p}{\partial x}\right) \tag{3.2}$$

Eqns (3.1) and (3.2) are the governing equations for the space-time evolution of the half film thickness $h(x,t)$ and can be solved together with the following boundary conditions. Symmetry requires $\partial p/\partial x = 0 = \partial h/\partial x$ at $x = 0$. At large $x$, we expect (see later) $p \sim x^{-4}$, which can be implemented as $\partial p/\partial x + 4p/x = 0$. By differentiating the geometric relation: $L(t) = h(x,t) + y_{in}(x,t)$ with respect to $t$ and using eqn (2.12) we have, at a suitably large value of $x = x_{max}$

$$\frac{\partial h(x_{max},t)}{\partial t} = \frac{dL(t)}{dt} + \left\{R_o\left(\frac{2 - 2\cos\theta_o - \theta_o \sin\theta_o}{\sin\theta_o - \theta_o \cos\theta_o}\right) - x_{max}\right\}\frac{df(t)}{dt} \tag{3.3}$$

The function $L(t)$ is specified to reflect how the drops are driven together or separated in an experiment. These boundary conditions together with a given initial shape of the drops for $h(x,t)$ at $t = 0$ allow eqns (3.1) and (3.2) to be solved numerically by the method of lines as a set of coupled differential-algebraic equations[19]. This completes the formulation of the interacting drop problem. The numerical solution will be independent of the precise magnitude of choice $x_{max}$ as long as it is sufficiently large. We will return to this point when we consider numerical results.

Eqn (3.1) can be integrated with the symmetry condition $\partial h/\partial x = 0$ at $x = 0$ and eqn (2.4) to give the outer asymptotic form at large $x$

$$h(x,t) \rightarrow h_{oo}(t) + \frac{x^2}{2R} + k(t) - f(t)\,x \tag{3.4}$$

with the function $h_{oo}(t)$ to be determined. We note that the linear and quadratic dependence of $h(x,t)$ at the outer edge of the film follow from the behaviour of the inner shape, $y_{in}(x,t)$ of the deformed drop and is a general consequence of simply integrating eqn (3.1).

## 4. Perturbation solution of film drainage equations

In this section we develop a perturbation solution of the Stokes-Reynolds film drainage equations given by eqns (3.1) and (3.2). The asymptotic behaviour in eqn (3.4) suggests we seek a perturbation solution of the form

$$h(x,t) = h_o(x,t) + h_1(x,t)$$
$$= [h_{oo}(t) + \frac{x^2}{2R}] + h_1(x,t) \quad (4.1)$$

Using the zeroth order solution: $h_o(x,t) \equiv h_{oo}(x,t) + x^2/(2R)$ in eqn (3.2) gives the zeroth order pressure relative to the pressure at infinity

$$p_o(x,t) = \frac{-3\mu \frac{dh_{oo}(t)}{dt} R}{2\left(h_{oo}(t) + \frac{x^2}{2R}\right)^2} . \quad (4.2)$$

This confirms the $p \sim x^{-4}$, $x \to \infty$ behaviour assumed in the previous section. Within this perturbation scheme, the scaled force $f_o(t)$ due to this pressure $p_o(x,t)$ is, according to eqn (2.5)

$$f_o(t) = -\frac{3\pi\mu R}{8\sigma} \frac{dh_{oo}(t)}{dt} \frac{[2R h_{oo}(t)]^{1/2}}{[h_{oo}(t)]^2} . \quad (4.3)$$

If we define the capillary number: $Ca = (\mu/\sigma)(dh_{oo}/dt)$ we see that the scaled force $f$ is proportional to a film capillary number: $Ca_f = Ca (R_o/h_{oo})^{3/2}$. In contrast, the corresponding film capillary number for interaction between axi-symmetric drops[13] is $Ca_f = Ca (R_o/h_{oo})^2$.

Inserting eqn (4.1) into eqn (3.1), the first order solution $h_1(x,t)$ satisfies

$$\sigma \frac{\partial^2 h_1}{\partial x^2} = -p_o(x,t) = \frac{3\mu \frac{dh_{oo}(t)}{dt} R}{2\left(h_{oo}(t) + \frac{x^2}{2R}\right)^2} . \quad (4.4)$$

This can be readily integrated with the symmetry condition $\partial h_1/\partial x = 0$ at $x = 0$ to give the complete perturbation solution

$$h(x,t) = \left[h_{oo}(t) + \frac{x^2}{2R}\right] + f_o(t) R_o \left\{\frac{2 - 2\cos\theta_o - \theta_o \sin\theta_o}{\sin\theta_o - \theta_o \cos\theta_o}\right\}$$

$$- \frac{2 f_o(t)}{\pi} \sqrt{2R h_{oo}(t)} \left\{\left(\frac{x}{\sqrt{2R h_{oo}(t)}}\right) \arctan\left(\frac{x}{\sqrt{2R h_{oo}(t)}}\right) + 1\right\}. \quad (4.5)$$

Here $h_{oo}(x,t) = L(t) - R_o(1 - \cos\theta_o)$ as we have used $L(t) = h(x,t) + y_{in}(x,t)$ and eqn (2.12) to simplify the result. We note that eqn (4.5) has the linear and quadratic behaviour at large $x$ as anticipated in eqn (3.4) as $[z \arctan(z)] \to (\pi z/2) - 1$ as $z \to \infty$.

This derivation of a perturbation solution follows the approach taken by Lai *et al.* but with two important differences. (i) The right hand side of eqn (4.4) has a different sign to that in eqn (2.6) of Lai *et al.* We believe this is an error that has propagated throughout their calculation. (ii) We impose the symmetry condition: $\partial h_1/\partial x = 0$ at $x = 0$ since the fluid interface must have continuous derivatives everywhere, whereas Lai *et al.* imposed the condition $h_1(x,t) \to 0$ as $x \to \infty$. Consequently, their solution of the fluid interface possesses an unphysical cusp at $x = 0$ and their function $h(x,t)$ does not have a linear term proportional to the scaled force $f$ at large $x$ as required by the general result in eqn (3.4). This linear term in $h(x,t)$ originates from the perturbation contribution $h_1(x,t)$ via the last term in eqn (4.5).

## 5. Numerical results and perturbation theory

We compare results from numerical solutions of the drainage and deformation equations of Hele-Shaw drops given by eqns (3.1) and (3.2) with the perturbation results from section 4. For simplicity we take the contact angle $\theta_o = \pi/2$ consistent with $x_1 = R_o$ (see Fig. 1). We calculate the deformation when the distance $L(t)$ is chosen to model cases where drops are driven together or driven apart from an initial half separation $h_{oo}(0)$ and accelerate smoothly from rest to a constant velocity. The form of $L(t)$ we use is:

$$L(t) = L_o + v_o (t - 1 + e^{-t/\tau}) \quad (5.1)$$

with the constant velocity parameter $v_o > 0$ ( or $< 0$) corresponding to the drops being driven apart (or driven together). We also consider the case where $L(t)$ is taken from experimental data[11,12] where the drops are first driven together and then separated.

To solve eqns (3.1) and (3.2) we choose the following scales[19]:

$$x_s \equiv Ca^{1/4} R_o, \qquad h_s \equiv Ca^{1/2} R_o,$$

$$t_s \equiv \mu Ca^{-1/2} R_o/\sigma, \qquad p_s \equiv \sigma/R_o, \qquad (5.2)$$

with $Ca \equiv \mu v_o/\sigma$ to non-dimensionalise eqs (3.1) and (3.2). We use the algorithm detailed in Carnie *et al.* with $(x_{max}/x_s) = 15$ and with 301 grid points between 0 and $(x_{max}/x_c)$ which is sufficient to give over 4 digits precision.

In Fig. 2 we compare numerical solutions of eqns (3.1) and (3.2) with the perturbation solution eqn (4.5) for two deformable drops with initial parabolic shape being separated from rest. All quantities have been scaled according to eqn (5.2). In Fig. 2a, we see that the central value at $x = 0$ of the deformation correction $h_1(0,t)$ is negative, that is, the drops deform towards each other as they are being separated. The rate of deformation and subsequent recovery predicted by the analytic perturbation theory, eqn (4.5), is too fast compared to the numerical solution. Interestingly, if we replace the analytic scaled force $f_o(t)$ in eqns (4.3) and (4.5) by the actual $f(t)$ obtained from the numerical solution, the result is in almost perfect agreement with the full numerical solution. The space-time evolution of the half film thickness $h(x,t)$ is given in the inset of Fig. 2a for the 4 time points A to D marked on the main graph. The apparent agreement between the numerical solution and the analytic perturbation result is due to the fact that the magnitude of the perturbation is small in this case.

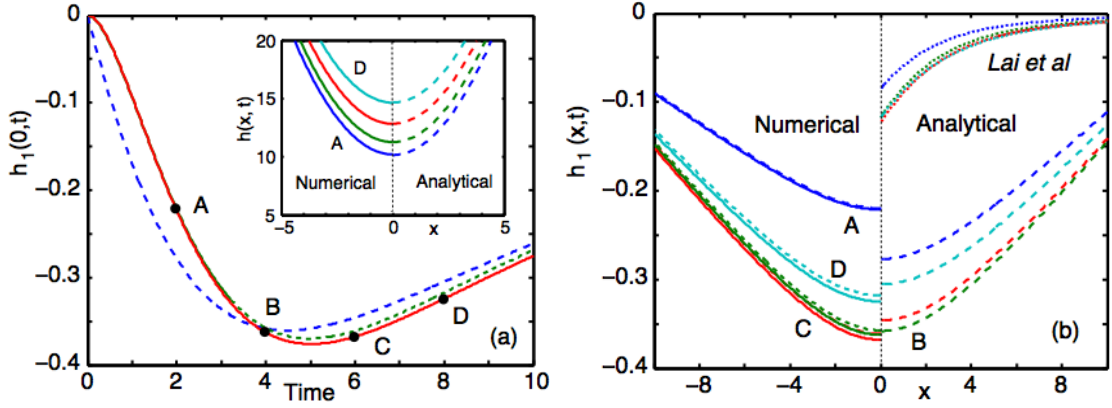

**Fig. 2** Evolution of the half film thickness $h(x,t)$ between two initially parabolic Hele-Shaw drops with initial scaled value $h_{oo}(t=0) = 10$, pulled apart with scaled velocity $dL(t)/dt = 1 - e^{-t}$ (cf eqn 5.2): numerical solution (—), analytic perturbation solution (eq 4.5) (- -) and analytic perturbation solution with the numerical force (•••) for (a) the central deformation $h_1(0,t)$ and half film thickness $h(x,t)$ in the inset, (b) perturbation deformation $h_1(x,t)$: left – numerical solution and analytic perturbation solution with the numerical force; right – analytic perturbation solution and corresponding results from the Lai *et al*.

In Fig 2b we see that the spatial form of the deformation $h_1(r,t)$ is not well predicted by the analytic perturbation theory for separating drops even though the magnitude of $h_1(x,t)$ ( < 0.4) is small compared to the half film thickness $h(x,t)$ (> 10). However, by replacing the analytic scaled force $f_o(t)$ in eqns (4.3) and (4.5) by the actual $f(t)$ obtained from the numerical solution, the deformation $h_1(x,t)$ can be predicted very accurately. This suggests that the analytic perturbation solution in eqn (4.5) has the correct functional form in the spatial coordinate $x$, but the scaled force is not well represented by the perturbation expression $f_o(t)$ in eqn (4.3).

In Fig. 3, we present the same set of comparisons for two parabolic drops initially at rest but driven together instead. In this case, the central deformation $h_1(0,t)$ is positive because the drops will flatten as they approach but the deformation predicted by the analytic perturbation solution is too large. The space-time evolution of the half film thickness $h(x,t)$ is given in the inset of Fig. 3a for the 4 time points A to D marked on the main graph. Note that at time D, the scaled central half thickness $h(0,t)$ has decreased from 10 to 5 where the scaled deformation $h_1(0,t)$ is about 1.5 which accounts for about 30% of the thickness. Note that by replacing the dimensionless force $f_o(t)$ in eqn (4.3) by the value of $f(t)$ obtained from the numerical solution, eqn (4.5) can give almost perfect agreement with the full numerical solution for the central deformation $h_1(0,t)$. This suggests that if $h_1(0,t)$ can be measured, eqn (4.5) can be used to infer the time-dependent scale force $f(t)$ quite accurately for approaching drops.

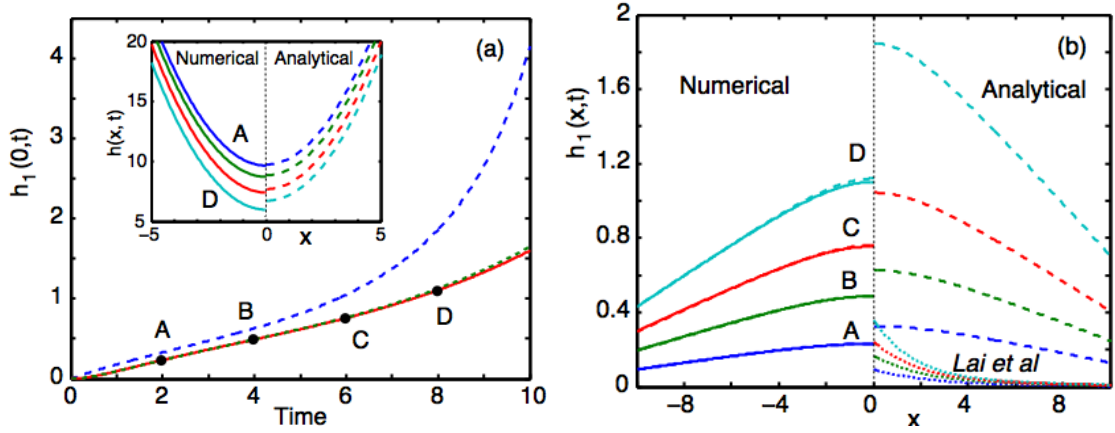

**Fig. 3** Evolution of the half film thickness $h(x,t)$ between two initially parabolic Hele-Shaw drops with initial scaled value $h_{oo}(t=0) = 10$, pushed together with scaled velocity $dL(t)/dt = -1 + e^{-t}$ (cf eqn 5.2): numerical solution (—), analytic perturbation solution (eq 4.5) (- -) and analytic perturbation solution with the numerical force (•••) for (a) the central deformation $h_1(0,t)$ and half film thickness $h(x,t)$ in the inset, (b) perturbation deformation $h_1(x,t)$ : left – numerical solution and analytic perturbation solution with the numerical force; right – analytic perturbation solution and corresponding results from the Lai *et al*.

In Fig 3b we see that the spatial form of the analytic perturbation theory for the deformation $h_1(r,t)$ is rather inaccurate. However, by replacing the analytic scaled force $f_o(t)$ in eqns (4.3) and (4.5) by the actual $f(t)$ obtained from the numerical solution, the deformation $h_1(x,t)$ then becomes almost indistinguishable from the full numerical result (left side of Fig 3b) which is in accord to the results seen in Fig. 3a.

We observe that the predictions according to the theory by Lai *et al.* for separating (Fig. 2b) and approaching (Fig. 3b) drops are qualitatively different from the results in the present calculation.

We now examine predictions of the Hele-Shaw model using parameters pertinent to the coalescence studies in microfluidic cells of Bremond *et al*. Two drops of radius $R_o$ = 30 μm are driven together in a microfluidic cell according to an experimentally controlled displacement function $d(t)$ between the centre of mass of the drops[12]. The experimental $d(t)$ values are fitted to a polynomial and differentiated to obtain the velocity as a function of time (Fig 4a). We see that the drops are initially at a centre-to-centre separation of 65 μm or an initial minimum separation of 5 μm. They are driven together initially until $t \sim 0$ and then separated (Fig. 4a). Using data corresponding to water drops in hexadecane ($\sigma$ = 50 mN/m, $\mu$ = 3 mPa s) the evolution of the half film thickness $h(x,t)$ on approach ($t < 0$) is given in Fig 4b corresponding to the time points A to E indicated in Fig 4a. We see clear evidence of flattening due to hydrodynamic interactions. As the separation is greater than 330 nm at all times, van der Waals interaction between the drops is negligible.

For $t > 0$, when the outer part of drops are being separated, corresponding to time points E to J, the central portion of the film continues to thin, see Fig 4c and attains the minimum thickness, at time J, of about 160 nm. Beyond this time, the central portion of the film also starts to separate as well, see Fig 4e.

A movie showing the space-time evolution of the film thickness that models this approach and separation experiment is available as on-line supplementary material.

To illustrate the history dependence and dynamic nature of drop-drop interaction we compare these results with the case of two stationary parabolic drops at an initial separation of 330 nm, the same separation at time E, $t = 0$, and separate them for $t > 0$ according to the same experimental velocity schedule as in Fig 4a. The evolution of this half film thickness $h(x,t)$ in this case is shown in Fig 4d. The result differs significantly from that for the approach-then-separate results in Fig 4c. The reason is that during the approach phase, the drops flattened considerably from the initial parabolic shape and the drop surfaces are still moving when the separation phase commences. Thus the behaviour on separation is very different from that of separating parabolic drops that are initially at rest. Consequently, any intuition based on studying the behaviour of separating parabolic drops from rest may be misleading when applied to moving drops[12]. We illustrate this point by comparing the central half separation $h(0,t)$ between these 2 cases in Fig 4e. We also show that the analytic perturbation theory for $h(0,t)$ for separating two initially stationary parabolic drops is also misleading as it predicts coalescence of the drops almost immediately upon separation (see arrow on Fig 4e). Although the maximum magnitude of the deformation predicted by the analytic perturbation theory for $h_1(0,t)$ is nearly correct (Fig 4f), this maximum deformation is attained too early in the separation phase and leads to the incorrect inference that the initially stationary parabolic drops may coalesce immediately upon separation.

## 6. Discussion

We have developed a model to describe the head-on dynamic interaction between two quasi two-dimensional drops in a Hele-Shaw cell. The model focuses on treating hydrodynamic interactions, surface deformation and thin film drainage in detail in the osculating region between the drops when the size of the deformation is small on the scale of the drop radius. Nonetheless, the effect due to global deformations arising from volume conservation needs to be included through matching of the outer drop shape to provide a physically consistent boundary condition for the differential equations that govern phenomenon in the inner thin film region.

While numerical solutions of the governing equations are required, we also gave an analytic perturbation solution. Even though this solution is not particularly accurate in some cases, it does provide useful physical insight into the dynamic properties in interacting drops for this system. The perturbation solution reveals that the amplitude of the deformations of the interface is characterised by the film capillary number: $Ca_f = (\mu v/\sigma)(R_o/h_o)^{3/2}$ and the profile of the thin film appears to diverge linearly at the outer edge (see eqn (4.5)). In contrast to a similar study of the dynamic interaction between three-dimensional axisymmetric drops[13,20] the corresponding film capillary number is: $Ca_f = (\mu v/\sigma)(R_o/h_o)^2$ and the profile of the thin film appears to diverge logarithmically at the outer edge. These apparently divergences are due to the mathematical form of the Young-Laplace equation in one-dimension or where there is axial symmetry, and these asymptotic forms are required to match to the global deformations of the drops via volume conservation of the drops.

On the other hand, accurate quantitative predictions of the evolution of the film thickness between drops can be obtained from the analytic perturbation solution in eqn (4.5) provided the correct numerical value of the dimensionless force $f(t)$ is used in place of the zeroth order perturbation expression $f_o(t)$, eqn (4.3). This means that if the film thickness can be measured, eqn (4.5) can be used to infer the time dependent force between moving drops in situations where the Hele-Shaw model is applicable.

In terms of the microfluidic experiments of Bremond *et al.*, the Hele-Shaw model does not appear to provide a good quantitative model. Using plausible physical parameters and the experimental drop separation function (see Fig. 4) we are not able to obtain films thinner than 160 nm. At this thickness destabilizing van der Waals forces are completely negligible. The reason for observing such thick films is that the Hele-Shaw model is equivalent to fluid drainage between two cylindrical drops for which the hydrodynamic repulsion between approaching drops is overestimated. Furthermore, the geometry of the experimental microfluidic cell is not very thin compared to observed drop radius. Indeed, the drops probably resemble oblate spheroids with an axis ratio of about 3:1. Using an axi-symmetric model[19] with a radius equivalent to the harmonic mean, the minimum separation can be as small as 5 nm where van der Waals forces can then be large enough to induce coalescence.

However, it is interesting to note that the Hele-Shaw model does predict the minimum separation would occur at around 1 ms which is within the spread of coalescence times observed in the microfluidic experiments[12]. Recall that we assumed the no-slip or immobile hydrodynamic boundary condition holds at the water/hexadecane drop interface. If on the other hand, the fully mobile or continuity of tangential stress condition were applied at the drop interface, the coalescence time would be smaller by more than an order of magnitude[21]. Our choice of the no-slip boundary condition at the fluid interface is motivated by the fact that when water is one of the fluids, trace amounts of surfactants or impurities will render the water interface to behave as a no-slip boundary[3,10]. The water/hexadecane viscosity ratio of 1/3 cannot account for the no-slip condition as the viscosity ratio must be much greater than $(R_o/h_o)^{1/2}$ (~ 15 in this case) in order for the drop interface to attain the no-slip condition[12].

**Acknowledgement**

We thank Dr Nicolas Bremond for discussions and for sharing a copy of his movie of coalescing drops in a microfluidic cell. This study was further stimulated by discussions between two of the authors (E. K. and R. M.) and Professor Howard Stone during his visit to the Institute of High Performance Computing in December 2008. This work was supported in part by the Australian Research Council, AMIRA International and State Governments of Victoria and South Australia. DYCC is an Adjunct Professor at the National University of Singapore.

**On-line supplementary material**

A MPEG4 movie of the result in Figure 4 is available on-line.

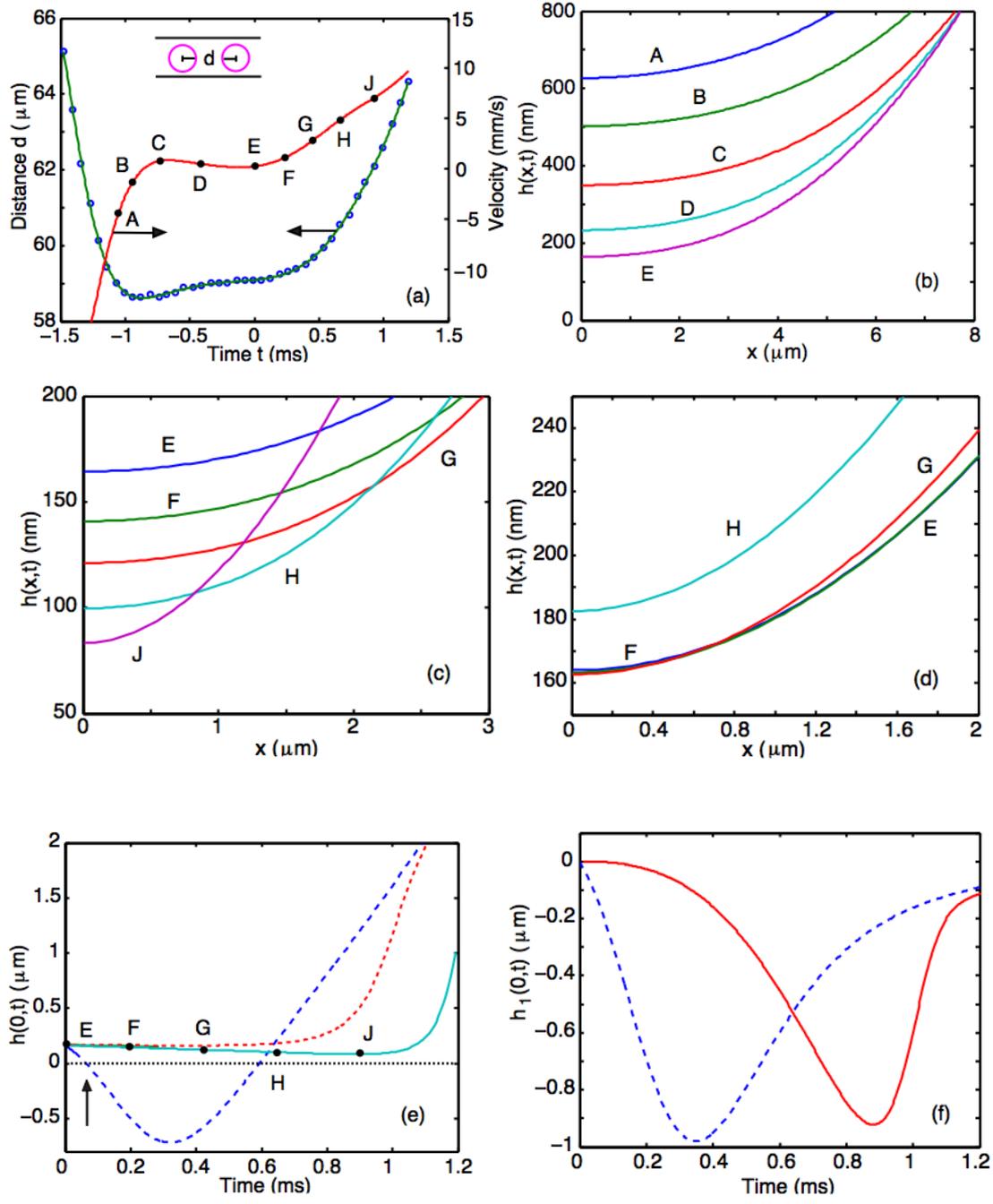

**Fig. 4** (a) The experimental separation $d(t) = 2 L(t)$ between the centre of mass of two interacting water drops ($R_o = 30$ μm) in hexadecane in a microfluidic cell: experimental data (o) from Lai et al[12] fitted with a 10th order polynomial (- -) (left ordinate) and the velocity $v(t)$ obtained by differentiation of the fitted polynomial (—) (right ordinate). (b) The hexadecane half film thickness $h(x,t)$ between the drops during approach at times marked in Fig. 4a. (c) The half film thickness $h(x,t)$ during separation at times marked in Fig. 4a. (d). The half film thickness $h(x,t)$ of an initially stationary parabolic film being separated at the same velocity as in Fig. 4c from the same minimum separation as the film at time E. (e) the central half film thickness $h(0,t)$: numerical solution of approach-then-separate (–•–), numerical solution of initially stationary parabolic drops (—) and analytic perturbation solution of initially stationary parabolic drops (- -). (f) central deformation $h_1(0,t)$: numerical solution of initially stationary parabolic drops (—) and analytic perturbation solution of initially stationary parabolic drops (- -). A movie of the approach-then-separate result is available as on-line supplementary material.

**Figure for Table of Contents**

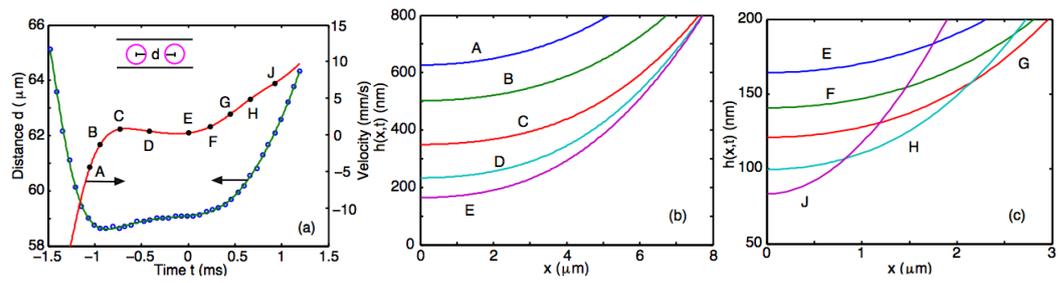